\documentstyle[12pt,aasms4]{article}
\newcommand{\doublespacing}{\let\CS=\@currsize\renewcommand{\baselinestretch}
{1.0}\tiny\CS}

\newcommand {\beq} {\begin{equation}}

\newcommand {\eeq} {\end{equation}}

\begin{document}

\title{Quantum Processes, Space-time Representation and Brain Dynamics}
\author{Sisir Roy\altaffilmark{1,2} and Menas Kafatos\altaffilmark{1}}
\centerline{$^1$Center for Earth Observing and Space Research, School of Computational Sciences}
\centerline{George Mason University, Fairfax, VA  22030  USA}
\centerline{$^2$Physics and Applied Mathematics Unit, Indian Statistical Institute, Calcutta 700108  INDIA}
\altaffiltext{1}{e-mail: mkafatos@gmu.edu;  sroy@scs.gmu.edu} 
\altaffiltext{2}{e-mail: sisir@isical.ac.in} 

\begin{abstract}
\noindent
The recent controversy of applicability of quantum formalism to brain dynamics
is critically analyzed. The prerequisites for applicability of any type of quantum formalism or quantum 
field theory is to investigate whether the anatomical structure of brain
permits any kind of smooth geometric Hilbert structure or four dimensional Minkowskian structure. The present 
understanding of brain function clearly denies any kind of space-time representation in the Minkowskian sense. However, three 
dimensional space and one dimensional time can be assigned to
the neuro-manifold and the concept of probabilistic geometry is shown to be an appropriate framework to understand 
the brain dynamics. The possibility of quantum structure for brain dynamics is also discussed in this framework.  
\end{abstract}
 \noindent

\noindent
\section{Introduction}
\noindent
The applicability of quantum formalism to brain dynamics has raised lot of interest among the scientific community 
(Tegmark 2000; Hagan et al.2002). Several authors (Stapp 1990;1993; Hameroff et al.1996; Riccardi et al.1967; Umezawa 1993; 
Alfinito et.al.2000; Jibu et al.1996) claimed that quantum processes and collapse
of wave function in the brain are of importance to help us to understand information processing and higher order cognitive 
activities of brain. Even before, Pribram (1991) had proposed his holographic model to help in understanding the information 
processing in the brain. However, the most fundamental issue which should be solved before applying any quantum 
approach has not been addressed by any of the above authors.
One of the prerequisites of applying any form of quantum mechanics to brain dynamics(in the non-relativistic domain) is to investigate 
whether the anatomical structure of brain permits assigning any kind of smooth geometric notion like distance 
function, or orthogonality relation between the vectors in the neuro-manifold. For applying quantum field theoretic 
model to memory function or spontaneous symmetry breaking, one needs to
construct space-time geometry in Minkowskian sense over this neuro-manifold. Then it is necessary to look into the  
plausibility of indeterminacy relation with Planck's constant $h$ or any other constant, say , a {\bf brain 
constant} at any level of brain functions. We think one should address these issues before applying any kind 
of quantum formalism to understand the information processing and higher order cognitive activities (Roy et al. 2003). \\ 
\indent
The plan of this paper is as follows:  At first we will analyze the anatomical structure of brain and its 
relation to Euclidean or non-Euclidean distance and then the possibility of assigning space-time (four dimensional) 
representation. Pellionisz and Llinas (1982; 1985) have shown that our present understanding of brain function 
does not permit to assign any space-time representation in the usual four dimensional form.
They considered tensor network theory where they assigned a metric tensor $g_{ij}$ to the Central Nervous System (CNS).
However, for global activities of the brain i.e.,to define the metric tensor over the whole neuro-manifold, this raises 
a lot of difficulties. For example, some cortical areas are non-linear or rough, so the tensor network theory becomes 
very much complicated and almost intractabale to solve the relevant mathematical equations. 
In one of our recent papers (Roy et al. 2002), we proposed that the statistical distance function may be considered over 
the entire neuro-manifold considering the selectivity properties of neurons(Hubel 1995). In this paper, we shall 
show that the statistical distance function and the statistical metric tensor aught to be considered, as they are 
expected to be very important concepts to understand the above mentioned issues. Then the possibility of Hilbert 
space structure and quantum processes are discussed with respect to structure of neuro-manifold.
\section{\bf Functional Geometry and Space-time Representation}
\indent
The internalization of external geometries into the Central Nervous System(CNS) and the reciprocal issue have created
lot of interest for the last two decades. The central tenet of the (Pellionisz and Llinas (1982) hypothesis is that brain 
is a tensorial system using vectorial
language. This hypothesis is based on the consideration of covariant sensory and contravariant vectors 
representing motor behaviour. Here, CNS acts as the metric tensor and determines the relationship 
between the contravariant and covariant vectors. The contravariant observable theorem has been 
discussed in the context of Minkowskian geometry as well as in stochastic space-time and quantum theory. It can be 
stated that measurements of dynamical variables are contravariant components of tensors.
This means that whenever a measurement can be reduced to a displacement in a particular coordinate system, it can be related to 
contravariant components of the coordinate system. To make an observation of a dynamical variable as position or momentum, 
the measurement is usually done in the form a reading of a meter or something similar to that. Through a series of calculations one 
can reduce the datum to a displacement of
a coordinate system. \\
\indent
Margenau (1959) analyzed this issue and claimed that the above reduction can give rise to a 
measurement if it satisfies the two requirements. It must be repeatable with the same results and must yield a  physically 
useful quantity. This can be easily shown in the context of Minkowski space. The motivation of Pellionisz and Llinas (1982) 
was to find a single underlying entity capable of representing any set of particular neuronal networks;  i.e., data derived 
from particular neuronal networks, generalized to the set of neuronal networks
(from a brain to the brain). In fact, if this is the case, it would imply that this is equivalent to consider the brain 
as a geometric object. However, they have shown that a space-time
representation (in the sense of Minkowskian  geometry) can not be assigned to the internal representation. The arguments can briefly be 
described as follows : \\ 
\indent
CNS can be compared with a procedure such as taking the picture of a moving object not with an instantaneous flash,  but replacing the
light with a set of axons (each having a different conduction time). Now through differently delayed neuronal signals, the simultaneous
external events will not be represented in the CNS as simultaneous. In other words, {\it the simultaneous onset of firings of a group of neurons 
with different conduction times will not produce a set of simultaneous external events}. Hence, the assignment of space-time 
geometry to the functional geometry of neurons is not possible, at least within the perview of our present understanding of brain dynamics.
It appears that a {\it three dimensional space and one time coordinate system} can be assigned to the internal representation. But to assign this kind of
space and time sturcture over the global activities of brain ( to account for, i.e., the holonomic like information processing)(Pribram 1991), one needs to
define a smooth metric tensor over the whole neuro-manifold. A family of neural networks forms a neuro-manifold. \\
\noindent
 However, as some cortical areas of brain are more {\it non-linear} and {\it rougher} than
others, it seems to be very difficult to construct a smooth metric tensor over the entire neuromanifold. In fact, the mammalian cerebral cortex 
has the form of a layered thin shell of grey matter surrounding white matter. The cortical mantle is one of the most important features 
of the brain and it plays a tremendously important role in understanding brain functions. Although the cortical surface is an important 
feature of mammalian brain, the precise geometry and variability of the cortical surface are little understood. Attempts have been made to
construct mathematical representation of a typical cortical surface using the data from the Van Essen Laboratory (Washington, Seattle, USA). 
This representation allows one to make statements about the geometry of the surface as well as its variability. Considering the surface as 
two-dimensional manifold in brain volume, it enables one to compute geometrical properties such as the {\it Mean} and the {\it Gaussian} 
curvature of the surface. However, much work is needed to understand the functional geometry of the brain. \\
\indent
Recently, Nakahara et al.(2002) used the concept of {\it Information Geometry} to understand the geometrical structure of 
a family of information systems.
 The family of information systems, stated here, are taken as consisting of a hierarchial structure of neuronal systems with feedback 
and feedforward connections. Amari introduced a duality structure in the Bayesian framework from the point of view of information geometry.
However, he considered a manifold equipped with Riemannian metric formed by a family of distributions. In this framework, the four dimensionai metric 
in the Minkwoskian sense is also not realizable due to the lack of existence of positive definite distribution functions for four dimensional space-time.
\section{\bf Quantum Formalism}
\noindent
Recent interests (Stapp 1993,1990; Hameroff et al.1996; Riccardi et al.1997; Umezawa 1993; Alfinito et al.2000; Beck 1996; Beck et al.1998) 
on the applicability of quantum formalism in understanding brain function leads one to consider several 
fundamental issues related to functional geometry of brain. In the quantum theory of mind-brain described by Stapp (1993;1990)
there are two separate processes: \\
 First, there is the unconscious mechanical brain process goverened by the Schr\"odinger equation.
It involves processing units that are represented by complex patterns of neural activity ( or more generally, of brain
activity) and subunits within these units that allow "association",i.e., each unit tends to be activated by the activation of 
several of its subunits. An appropriately described mechanical brain evolves by the dynamical interplay of these associative units. 
Each quasi-classical 
element of the ensemble that constitutes the brain creates, on the basis of clues, or cues, coming from
various sources, a plan for a possible coherent course of action. Quantum uncertainties entail that a host of different possibilities 
will emerge. This mechanical phase of the processing already involves some selectivity, because the various input clues contribute 
either more or less to the emergent brain process according to the degree to which these inputs activate, via associations,
the patterns that survive and turn into a plan of action. \\
\indent
Hameroff and Penrose (1996) discussed the issue of quantum coherence and consciousness in following way :
\begin{enumerate}
\item Quantum coherence and the wave function collapse are essential for 
consciousness and occur in cytoskeleton microtubules and other structures within each of the brain's neurons.
\item Quantum coherence occurs among tubulins in micro-tubules, pumped by thermal and 
biochemical energies. Evidence for some kind of coherent excitation in proteins has 
recently been reported by Vos et al.(1993). The feasibility of quantum coherence in, seemingly, noisy, chaotic cell 
environment is supported by the observation that quantum spins from biochemical radical pairs which become separated, retain their
correlation in the cytoplasm.
\item  For the Orchestrated Objective Reduction (OrchOR), as put forward by Penrose and Hameroff, each of superposed
states have their own space-time geometries. When the degree of coherent mass-energy difference
leads to sufficient separation of space-time geometry, the system must choose and decay to a single
universe state. Thus the OrchOR model makes it necessary for inclusion of self selections 
in the fundamental space-time geometry.
\item It is known that a typical brain neuron has roughly $10^7$ tubulins. If the tubulins in each neuron are involved in quantum 
coherent state, then roughly $10^3$ neurons would be required to sustain coherence for 500 msec, by which time, according to 
Penrose and Hameroff, the quantum gravity threshold is reached and only then OOR occurs.
\end{enumerate}
According to these authors, one possible scenario for emergence of quantum coherence leading to OrchOR and 
conscious events is cellular vision. Albrecht-Buehler (1992) has observed that single cells 
utilize their cytoskeletons in cellular vision-detection, orientation and directional response to beams of 
red/infrared light. Jibu et al.(1996) argued that this process requires quantum coherence in micro tubules and ordered water. \\
\indent
Pribram (1991) developed the first holographic model of the brain. According to
Pribram, the brain encodes information on a three dimensional energy field that
enfolds time and space, yet allows to recall or reconstruct specific images from
the countless millions of images, stored in a space slightly smaller than a melon. Further, he and his collaborators studied the 
implications of Gabor's quanta of information (1946) on brain function and their relation to Shannon's measure (1948) for the amount
of information related to the data, as obtained in their investigations. Furthermore,  Pribram along with Jibu and Yasue(1996) investigated how the 
quantum mechanical processes can operate at the synaptodendritic level. According to their view, something
like the phenomenon of superconductivity can occur by virtue of boson condensation over short ranges when the water molecules adjacent 
to the internal and external hydrophilic layers of the dendritic membrane become aligned by the passive
conduction of post synaptic excitatory and inhibitory potential changes, initiated
at synapses (Jibu et al. 1996). It is generally argued that the brain is warm and wet. It is interesting to note that recent theoretical 
and experimental papers support the prevailing opinion  (Tegmark 2000) that such large warm systems will rapidly lose quantum
coherence and classical properties will emerge. In fact, the rapid loss of coherence
would naturally be expected to block any crucial role for quantum theory in explaining the interaction between our conscious 
experiences and the physical activities of our brains. However, very recently, Hagan et al.(2002) claimed that at certain 
level the quantum coherence can be retained.\\
\indent
It is clear from the above discussions that it is necessary to use either non-relativistic quantum
mechanics or quantum field theory for the description of activities of brain.
For any kind of quantum field theoretic approach one needs to define the field functions over relativistic space-time as a prerequisite.
So far, experimental evidence regarding the anatomy of brain does not permit the space-time description (in a Minkowskian
sense), and consequently, any type of quantum field theoretic model is hard to be conceivable, at least, at the 
present state of understanding in brain dynamics.\\
However, in order to apply any kind of non-relativistic quantum mechanics, the basic prerequisites are Hilbert
space structure and the indeterminacy relation. Before going into the context
of indeterminacy relation in brain dynamics, let us try to analyze the micro- and macro- structures in the brain.
\subsection{\bf Brain activity at various scales}
\noindent
The different scales of activities of brain can be classified in the following
manner (Freeman 1999):
\begin{enumerate}
\item Pulses of single neurons, microtubles in milliseconds and microns may be considered as the part of microstructures.
\item Domains of high metabolic demand managed in seconds and centimetres (for measuring the spatial patterns of cerebral blood flow). This 
can be designated as the macrostructure.
\item Millimeters and tenths of a second are the patterns of the massed dendritic potentials in EEG recordings from the waking and 
sleeping brains. This can be considered as an area where there might be a level of mesostructure.
\end{enumerate}
Freeman (1991) suggested that the perception cannot be understood solely by examining properties of individual 
neurons, i.e., only  by using the microscopic approach (dominant approach in neuroscience). Perception depends 
on the simultaneous activity of millions of neurons spread throughout the cortex. Such global activity has to be of the macroscopic approach. \\
\indent
Micro- or macrostructures in the brain are distinguished by the scale of time or energy. The macrostructure can be characterized 
by the fact that brain lives in hot and wet surroundings with $ T = 300 K$. This should be discussed in the context of quantum coherence 
versus thermal fluctuations. The physiological temperature $T = 300 K$ corresponds to an energy $ E \sim 1.3 \times 10^{-2}$ eV.
Let us now define a signal time $\tau = 2 \pi \omega$ and $\hbar\omega = E $.
As $\hbar\omega =E $, we can get
$$ \omega = 2 \times 10^{13} s^{-1} \ \ \ {\rm or} \ \ \ \tau = .03 {\rm ps}$$
\noindent
This indicates that the  physiological temperatures correspond to frequencies {\bf smaller} than the {\it picosecond} scale. 
They correspond to the time scale involving electronic transitions like electron transfer or changes in molecular bonds.
In cellular dynamics, the relevant time scale is of the order of \ $\tau > 0.4$ ns, \ where, \ $E_{\rm cell} \sim 10^{-5}$eV \ (Beck 1996).
To allow comparison with quantum scales, let us distinguish the two scales as
follows:
\begin{enumerate}
\item  The macroscopic or celluar dynamics level with time scales in the {\it milli}, down to the {\it nanosecond} range.
\item The microscopic or quantal dynamics level with time scales in the {\it pico}, down to the {\it femtosecond} range.
\end {enumerate}
The large difference between the two time scales indicates quantum processes might be involved in the individual microsites and decoupled from the 
neural networks. Recently, using dimensional analysis, Bernroider(2003) made an analysis by applying the concept of Lagrangian action to brain
processes at different scales of resolutions in order to clarify the current dispute whether
classical neurophysics or quantum physics are relevant to brain function. \\
\indent
The central issue of his analysis is to estimate the order of action from dimensional analysis, 
relevant to brain physiology and its close proximity to quantum action i.e., $\hbar$ (Planck's constant).
For example, spiking action at the single cell level is found to involve $1.8\times 10^{-15}$
Lagrange (using mechanical units) down to $2.1\times 10^{-16}$. This lies between $10^{18}$
and $10^{19}$ times Planck's constant. This is in good agreement with the time scale
difference and spiking time estimated by Tegmark (2000). However, he pointed out that the action behind
the selective ion permeation and channel gating might be of interest at the molecular level.
Considering $10^5$ ions permeating per msec and employing a saturating barrier model with
one or two ions crossing during that time, the action turns out to be of the order of $0.48\times 10^{34}$ Lagrange,
which is in the range of the quantum action $\hbar$ ($1.05459\times 10^{-34}$  MKS units).
It implies that brain functioning at a certain level might be a proper arena to apply the quantum
formalism. 
\subsection{\bf Indeterminacy Relations}
\noindent
In communication theory, Gabor (1946) considered
an uncertainty relation between frequency($\omega$) and time($t$)as
$$\delta \omega \delta t = \frac{1}{2}$$
This is similar to the Heisenberg energy($E$)-time($t$) uncertainty relation :
$$\delta E \delta t = \hbar \ \ \ \ \ \  [\hbar=\frac{h}{2\pi}]$$
where $h$ is Planck's constant.
Now if the quantum formalism is valid (even in its non-relativistic form) in brain dynamics, there should
exist a similar type of uncertainty relation between frequency/energy and time, i.e.,
$$\delta\omega \delta t = b $$
where  $b$ may be termed here as the ``brain constant''. Even if there exists such a constant in brain dynamics, one needs to 
relate it to action quanta like Planck'sconstant. {\it Future research in brain function can shed light on this important issue}. \\
\indent
As far as the existence of Hilbert space structure is concerned, one needs to define a smooth distance function
over the cortical surface of the brain. It should be mentioned that Joliot et al.(1994) found
a minimum interval in sensory discrimination. Considering this aspect, they claimed that consciousness is a non continuous event
determined by the activity in the thalamocortical system. Now, if this is the case, then one needs to introduce discrete 
time or some form of granularity in space and time geometry. 
\section{ Probabilistic Geometry and the Neuromanifold}
\noindent
Let us now describe first the geometroneuro-dynamics as proposed by Roy and Kafatos (2002) considering
the  neurophysiological point of view. Then we shall generalize the approach and investigate the relevant issues 
from a more generalized perspectives.
\subsection{\bf Orientation Selectivity of Neurons and Statistical Distance}
\noindent
Recent research on Planck scale physics (Roy 2003a) sheds
new light on the possible geometrical structure for discrete and continuum levels.
The idea of probability in geometric structure as proposed and developed by Karl Menger around 1940, (1942;1949)
seems to be a very useful tool in defining distance function over the cortical areas of brain. 
\indent
There is a  large variety as well as 
number of neurons in the brain. Collective effects which can only be accounted for in terms
of statistical considerations, are  clearly important in such case. Experimental evidences
point to  more than 100 different type of neurons in the brain, although
the exact number is not yet decided. It is found that no two neurons are
identical, and it becomes very difficult to say whether any particular
difference represents  more than the other i.e., between  individuals or between different classes.
Neurons are often organized in clusters containing the same type of cell. The
brain contains thousands of clusters of cell structures  which may take the
form of irregular clusters or of layered plates. One such example is the
cerebral cortex which forms  a plate of cells with a thickness of a few
millimeters.  \\
\indent
In the visual cortex itself (Hubel 1995), certain
clear, unambiguous patterns in the arrangement of cells with particular 
responses have been found. Even though our approach could apply to non-visual neurons, here
we limit our study to the neurons in the visual cortex as the visual cortex
is smoother preventing non-linear effects. For example, as the measurement 
electrode is moving at right angles to the surface through the grey matter, 
cells encountered one after the other have the same orientation as their receptive field axis. It is also found 
that if the electrode penetrates at an angle, the axis of the receptive field 
orientation would slowly change as the tip of the electrode is moved through 
the cortex. From a large series of experiments in cats and monkeys it was 
found : \\
\indent
{\it Neurons with similar receptive field axis orientation are located on 
top of each other in discrete columns, while we have a continuous change of 
the receptive field axis orientation as we move into adjacent columns}. \\
\indent
The visual cortex can be divided into several areas. The most important areas in the visual cortex  are  V1,V2,V3,V4 and MT (V5).
The primary visual cortex area V1 is important for vision. It is the principal entry point for visual input to the cerebral cortex. 
From this area, relays pass through a series of visual association areas in parietal and temporal regions and finally to the prefrontal cortex 
where substrates for decision making on the basis of visual cues are found. The main issues related to visual cortex are linked to intrinsic and
extrinsic relays of each cortical region, geometrically ordered microcircuitry 
within appropriate areas etc. Because of the stripy appearance of area V1, this area is
also known as the striate cortex. Other areas of visual cortex are known as the extrastriate (nonstriate) visual cortex. \\
\indent
In the monkey striate cortex, about 70\% to 80\% of cells have the
property of orientation specificity. In a cat, all cortical cells seem to be
orientation selective, even those with direct geniculate input (Hubel,1995).
Hubel and Wiesel found a  striking difference among orientation-specific
cells, not just in the optimum stimulus orientation or in the position of 
the receptive field on the retina, but also in the way cells behave. 
The most useful distinction is between two classes of cells : simple and complex. These two types differ
in the complexity of their behavior and one can make the resonable assumption
that the cells with the simpler behavior are closer in the circuit to the input of the cortex. \\
\indent
The first oriented cell recorded by Hubel and Wiesel(1995)  which responded to the
 edge of the glass slide was a complex cell. The complex cells seem to have
 larger receptive fields than simple cells, although the size varies. Both type of cells
 do respond to the orientation specificity. There are certain other cells which respond not only to the
 orientation and to the direction of movement of the stimulus but also to
 the particular features such as length, width, angles etc. Hubel and Weisel
 originally characterized these as hypercomplex cells but it is not clear
 whether they constitute a separate class or, represent a spectrum of more or less complicated receptive fields. 
We now ask how the computational structure or filters can manifest as orientation detectors? \\
\indent
 Pribram(1981) discussed this question, whether single
neurons  serve as feature or channel detectors. In fact, Pribram and his
collaborators( 1981 \& references therein) made various attempts to classify "cells" in the
visual cortex. This proved to be impossible because each cortical cell 
respondedto several features of the input such as orientation, velocity, the
spatial and temporal frequency of the drifted gratings. Furthermore, cells and cell
groups displayed different conjunctions of selectivities. \\
\indent
From these findings and analysis, he concluded
that cells are not detectors, that their receptive field properties could be
specified but that the cells are multidimensional in their characteristics
(Pribram 1991). Thus, the pattern generated by an ensemble of neurons is 
required to encode any specific feature, as found  by Vernon Mountcastle's
work on the parietal cortex and Georgopoulos's data(Pribram 1998) on 
the motor cortex. \\
\indent
It is worth mentioning that Freeman and his collaborators(1991) suggested
that perception cannot be understood solely by examining properties
of individual neurons i.e. by using the  microscopic approach that currently dominates
neuroscience research. They claimed that perception depends on the
simultaneous, cooperative activity of millions of neurons spread throughout
expanses of the cortex. Such global activity can be identified, measured and
explained only if one adopts a macroscopic view alongside the microscopic
one. 
\subsection{\bf Statistical Distance}
\noindent
We can define the notion of distance between the ``filters'' or the orientation
selective neurons. This distance is similar to the statistical distance
between quantum preparations as introduced by Wotters (1981). The statistical
distance is most easily understood in terms of photons and polarizing filters : \\
\indent
Let us consider a beam of photons prepared by a polarizing filter and analyzed
by a nicol prism. Let $\theta \in [0,\pi]$ be the angle by which the filter
has been rotated around the axis of the beam, starting from a standard
position ($\theta =0$) referring to the filter's preferred axis as being vertical.
Each photon, when it encounters the nicol prism, has exactly two options : to
pass straight through the prism (with  ``yes'' outcome) or to be deflected in a specific direction 
characteristic of the prism ( ``no'' outcome). Let us assume that the orientation 
of the nicol prism is fixed once and for all in such a way that vertically 
polarized photons always pass straight through. By counting how many photons yield each of the two possible 
outcomes, an experimenter can learn something about the value of \ $\theta$ \ via the formula 
\ $ p= cos^2 \theta$, where $p$ is the probability of ``yes''(Wotters 1981), as
given by quantum theory. \\
\indent
If we follow this analogy in the case of oriented neurons in the brain i.e. as
if the filters are oriented in different directions like oriented analyzers,
we can proceed to define the statistical distance.
\subsection{\bf Statistical Distance and Hilbert Space}
\noindent
It can be shown (Wotters 1981) that the statistical distance between two preparations
is equal to the angle in Hilbert space between the corresponding rays. The main
idea is as follows: \\
\indent
 Let us imagine the following experimental set up. There are
two preparing devices, one of which prepares the system in a specific state, say
$\psi^1$, and the other prepares in $\psi^2$. Here, the statistical distance
between these two states can be thought as the measure of the number of distinguishable
preparations between $\psi^1 \& \psi^2$. However, in treating quantum systems,
new features should be observed as opposed to rolling the dice. For dice,
there is only one possible experiment to perform, i.e., rolling the dice, whereas
for a quantum system, there are many, one for each different analyzing device.
Furthermore, two preparations may be more easily distinguished with one analyzing
device than with another. For example, the vertical and horizontal polarizations of 
photons can easily be distinguished with an appropriately oriented nicol prism, but can not
be distinguished at all with a device whose eigenstates are the right and left handed circular 
polarizations. Due to this reason, one can speak of the statistical distance between
two preparations $\psi^1 \& \psi^2$ as related to a particular measuring
device which means the statistical distance is device dependent. The absolute statistical
distance between $\psi^1 \& \psi^2$ is then defined as the largest
possible such distance, i.e., statistical distance between $\psi^1 \& \psi^2$
when the states are anlyzed by the most appropriate or discriminating apparatus. We can 
illustrate this point in the following way: \\
\indent
 Let $\phi_1, \ \phi_2$ ..... $\phi_N$ be the eigenstates of a measuring device $A$, by which $\psi^1 \& \psi^2$
are to be distinguished. It is assumed that these eigenstates are non-degenerate
so that there are N-distinct outcomes of each measurement. The probabilities
of various outcomes are $|(\phi_i,\psi^1|^2$ if the apparatus is described by $\psi^1$ and
$|(\phi_i,\psi^1|^2$ if the apparatus is described by $\psi^2$. Then the statistical distance
between $\psi^1 \& \psi^2$ with respect to the analyzing device $A$ is
\begin{equation}
d_A(\psi^1,\psi^2) = cos^{-1} [\Sigma_{i=1}|(\phi_i,\psi^1)||(\phi_i,\psi^2)|]
\end{equation}
This quantity attains its maximum value if it assumes one of the eigenstates of $A$, 
(say, $\phi_1$). In that case, we get the statistical distance as
\begin{equation}
d(\phi^1,\phi^2) = cos^{-1}|\phi^1,\phi^2|
\end{equation}
\noindent
This clearly indicates that the statistical distance between two preparations is
equal to the angle in Hilbert space between the corresponding rays. The equivalence
between the statistical distance and the Hilbert space distance might be very
surprising at first. It gives rise to the interesting possibility that statistical
fluctuations in the outcome of mesurements might be partly responsible for Hilbert space
structure of quantum mechanics. These statistical fluctuations are as basic as the 
fact that quantum measurements are probabilistic in their nature. \\
\indent
However, it should be mentioned that although representation of orientation of
objects in the visual cortex is fairly fine-scaled, visual information regarding
the nonstriate visual processing and in superior colliculus is very rough and
varies in a non-linear way from that in striate cortex. This type of nonlinearity
is neglected here as we have considered statistical considerations which average
out this type of nonlinearity. Instead, we considered here the distance between the
different clusters of neurons or between the ensemble of neurons.
\subsection{Perception and Relational Aspects in Probabilistic Geometry}
\noindent
The issue of continuum and discreteness remains a long standing problem over the last few centuries.
In mathematics, if the quantity $A$ is equal to the quantity $B$ and $B$ is equal to $C$, then $A$ is
equal to $C$, i.e, mathematical equality is a transitive relation. In the observable continuum ``equal''
means indistinguishable. In psychology, following Weber and Fechner (1860), we can say that $A$ may lie within the
threshold of $B$ and $B$ within the threshold of $C$. Poincar\'e (1905) suggested  \\
``{\it for the raw result of experience,\ \ $A = B, B = C; \ \ \  \ A \le C$ \ \ , which may be regarded as the formula 
for the physical continuum}'' \\
\noindent
that is to say that physical continuum is not a transitive relation. Menger (1949) tried to solve this problem  
from the positivist point of view. Following his words : \\
\noindent
`` {\it Instead of distinguishing between a transitive mathematical and intransitive physical relation, it thus seems more hopeful to
retain the transitive relation in mathematics and to introduce for the distinction of physical and psychological quantities a
probability, that is, a number lying between 0 and 1}''. \\     
\indent
He considered the role of probability in geometry and introduced the concept of probabilistic metric.
He introduced the concept of a set of hazy lumps instead of considering set of points. Then the problem turns out to be similar to finding 
a probability of the overlapping lumps. For more intuitive understanding, the lumps were considered as the ``seat''
of elementary particles like electrons, protons etc. These lumps are taken as not to be reducible to any other structures.
In other words,they are the ultimate building blocks of space and time. Therefore, a kind of {\it granularity} is introduced here at the very
basic level. Mathematically speaking, it can be stated as : \\
\noindent
`` {\it for each pair of elements $A$ and $B$ of probabilistic geometry, it is possible to associate a distribution
function $F_{AB}(z)$ which can be interpreted as the probability that the distance between the points is less than
$z$}''. \\
\indent
Essentially the relational aspect of geometry has been proposed and elaborated by Menger.
In our approach, towards geometroneurodynamics ( Roy and Kafatos 2002), we have considered the same relational aspect
of geometry by considering the orientation selectivity of neurons. Recently, Mogi(1997) tried to
reinterpret Mach's principle in the context of the response selectivity of neurons.
He proposed that in perception, the significance of firing of a neuron is determined by the 
relation of the firing to other neurons at that very psychological moment. He called it {\it Mach's principle
in perception}. According to his proposal it is not meaningful to talk about the firing
of a single neuron in isolation and its significance in perception. \\
\indent
Menger replaced the usual metric function by a distribution function and showed that this distribution function 
satisfies all the axioms of the metric. Hence it is known as {\bf probabilistic metric space}. 
There are various types of probabilistic metric spaces used in differenti branches of physical science 
(Schewizer and Sklar 1983; Roy 1998). On a large scale, taking averages over these distributions,
one can get the usual metric structure. \\
\indent 
Here, we shall consider the stochastic space as proposed by Frederick(1976). 
In this model, the actual points of the space are stochastic in nature. These points can not be used as either a basis for
a coordinate system or to define a derivative. However, the space of common experience at large scales or in the laboratory frame is
nonstochastic. Therefore, we can start from large scale nonstochastic space and  continue mathematically
to stochastic space i.e. towards small scales. This stochasticity is cosnsidered to be manifested in a stochastic metric
$g_{ij}$ and the corresponding mass distribution determines not only the space geometry but also the space stochasticity. 
However, as more and more mass is confined in a region of space, the less stochastic will be that space.
 The relation between covariant and contravariant quantity can be written as 
$$x^i = g^{ij}x_j$$ 
As $g_{ij}$ is stochastic, one obtains a distribution of the contravariant quantity $x^i$ instead of a fixed quantity.
This may play a significant role in accounting for the tremor in  motor behaviour in neurophysiological experiments which will be considered in 
future publications.
Now let us assume that a Lagrangian can be defined taking this stochastic metric. Then we can define a pair of conjugate variables, 
one covariant and another contravariant, as
$$P_j = \frac{\partial L}{\partial {\dot{q}}^j}$$
where $P_j$ \ is a covariant quantity. However, $P_j$ is not observable in the laboratory as it is covariant. The observable 
quantity is the covariant one , i.e.,
$$P^j = g^{j\nu} P_\nu$$
As the metric is stoachastic in nature, $P^j$ is a random variable. So, if one member of the conjugate pair is well defined, the other 
member will be random.
\subsection{Neurophysiological Basis for Stochasticity in Metric}
\noindent
 Let us now look into the origin of stochasticity in neuromanifold.
The neurophysiological evidences shows that most neurons are spontaneously active, spiking at random intervals in the
absence of input. Different neuron types have different characteristic spontaneous rates, ranging from a few spikes
per second to about 50 pikes per second. The mechanism of regular activity is well studied whereas the mechanism of
random spontaneous activity is now not well understood. Several possibilites are discussed by Lindhal and Arhem (1994). \\
\indent
One is the well known ion-channel hypothesis. According to this, the nerve impulses are triggered by the opening
of single ion channels where the ion channel gating is random. Ion channels are membrane proteins through which the
current causing the nerve impulse passes. Donald (1990) considered that the randomness may be related to quantum fluctuations.
Lindhal et al. (1994) suggested that single channels may cause spontaneous activity in areas of 
the brain with consciousness. However, the detailed mechanism of ion-channel gating is still not well understood. 
Grandpierre (1999) made an attempt to study the effect of the fluctuation of the zero point field(ZPF) in the
activity of brain. As such, the future investigations on the effect of ZPF on the neurons may shed new insights not
only for the spontaneous activity of neurons but also on the actual process of consciousness.
We like to emphasize that in our picture, the fluctuation associated with this kind of spontaneous activity of neurons is the
physical cause behind the stochasticity of metric tensor.
\noindent
To start with, let us take Frederick's (1976) version of stochasticity in geometry. He made several interesting postulates 
as follows: 
\begin{enumerate}
\item {\bf The metric probability postulate} :
$P(x,t) = A \sqrt(-g)$ , where for a one-particle system  $P(x,t)$ is the particle probability distribution, $A$ is a real
valued function, and $g$ is the determinant of the metric.
\item {\bf The metric supersposition postulate} :
If at a position of a particle the metric due to a specific physical situation is $g_{ij}^1$ and the metric
due to a different physical situation is $g_{ij}^2$ , then the metric at the position of the particle due to the presence
of both of the physical situations is $g_{ij}^3$ can be written as
$g_{ij}^3 = \frac{1}{2} [ g_{ij}^1 + g_{ij}^2]$
\item {\bf The metric $\psi$ postulate} :
There exists a local complex diagonal coordinate system in which a component of the metric is at the location of the
particle described by the wave function $\psi$.
\end{enumerate}
\noindent
 We have started with Frederick's approach not only for the attarctive mathematical framework
for neuromanifold but also for the use of Mach's principle as the guiding rule for stochastic geometry.
It then becomes possible to derive quantum mechanics by adopting a strong version of Mach's principle
such that in the absence of mass, space becomes non-flat and stochastic in nature. At this stage, the stochastic metric
assumption is sufficient to generate the spread of wave function in empty space. Thus, following this 
framework, one obtains an uncertainty product for contravariant position vector ($q^1$) and cantravariant
momentum vector ($P^1$)as
$$\Delta q^1 \Delta P^1 =  \Delta q^1 \Delta (P_\nu g^{\nu 1})$$
where $P^j = g^{j\nu} P_\nu$. Now the question is what is the minimum value of this product. It can be shown that
$$\Delta q^1 \Delta(P_{\nu_{\rm min}} g^\nu 1) = \Delta q^1 \Delta P^1  > k_{\rm min}$$
which is nothing but the uncertainty principle  with $k$ as the action quantity similar to  $h$, i.e., Planck's constant. 
Moreover, using the superposition postulate of metric tensor, it is also possible to account for the interference phenomena.
\subsection{\bf Penrose and Hameroff Approach : OrchOR Model}
\noindent       
It is to be noted that the above metric supersposition postulate can be shown to be valid under the weak approximation
of the general theory of relativity. If there is more non-linearity in the cortical surface, the superposition
may actually break down. In Penrose and Hameroff (Hameroff 1996) model, they considered a kind of superpostion of space and time
geometries in order to relate it with the superposition of wave functions and the decoherence due to variation of
mass distribution and hence due to gravity effects. In the above framework,  it is possible to  relate
the superposition of wave functions using the superposition postulate of metrics. Becaues of the existence
of different curvatures at different points ( due to different mass distributions) in the framework of statistical
geometry, one can write the superposition of metrics or geometries. But it is necessary to investigate the real neurophysiological
conditions under which the superposition of metrics would be a valid approximation.
It raises a new possibility of constructing a Hilbert structure over the neuromanifold within the framework
of statistical geometry. The construction of Hilbert structure over the neuromanifold is one of the prerquisites
for applying any kind of quantum mechanical process to the brain.
\section{ Information Processing in the Brain}
\noindent
The information generated by integrated neural processes and its measurement has created a lot of interest among 
the scientific community for the last few years. The measure of information essentially depends on the basis of 
statistical foundation of information theory (Shannon 1948). One of the intriguing question arises is how far the statistical
 aspects of information theory can help one to assign  a measure to differentiate the informative character of the
 neural processes without any reference to an external observer .
The issue of the external observer has been debated in various branches of science and philosophy over the last century 
since the birth of quantum mechanics. In fact, the issue of measurement procedure in the history of science has been 
reanalyzed and still under active consideration after the mathematical formulation of Von Neumann using the 
statistical concept of entropy.  In the standard approach, one generally assigns a number to measure the information 
and probability of the states of the system that are distinguishable from the point of view of an external observer. 
 But the brain not only processes the information but also interprets the pattern of activities (Pribram 1991). Therefore, one must 
avoid the concept of a privileged viewpoint of an external observer to understand the information processing in the 
neural processes in the brain. \\
\indent
In our approach, we have developed a framework (Roy et al. 2003b) where it is possible to
 avoid the concept of an external observer by reanalyzing the very basis of measurement procedure as well as the 
neurophysiological evidences in the standard paradigm. Edelman et al.(2000) discussed this problem in the context of 
neurophysiology and consciousness. The main problem is 
how to measure the differences  within a system like the brain ?
He defined a concept of mutual information for this purpose. Here, they considered the entropic measure to define the 
information as considered in Shannon's framework. \\
\indent
The principal idea that lies behind our approach can be summarized as follows : \\
The concept of invariance plays a crucial role to understand the {\it information processing} and {\it measurement issues
} in the brain. In the brain, a matching occurs between an input pattern and a pattern inherent in the synaptodentritic 
network by virtue of generic or learning experience. In the Holonomic theory, both the input and output patterns provide the 
 initial conditions. The match between them is considered to be probabilistic in nature (Pribram 1991). We have introduced 
here a kind of invariance assisted by the context (as described by the inherent patterns in 
the dendrites). This is quite similar to environment-assisted invariance in Quantum Mechanics (Zurek 2002).  
It is one of the fundamental principles of quantum mechanics known as {\it quantum determinism} where it is possible to show that an 
entangled  state is formed between the input pattern and the output pattern. This state can be written  in terms of 
its basis vectors. Now picking up the specific term from the expansion is  generally known as selection. Redhead (1989)  
emphasized that the selection of the parts is related to the attention to particular subensemble of the whole. This 
means selection is not part of   quantum physics.  \\
\indent
In physicist's language, the selection signifies measurement that 
marks the end of quantum physics. In contrast, the ``yes-no''- experiment  puts the selection process at the beginning and 
makes the involvement of brain dynamics (or  the selection that underlies the pattern recognition in the brain) into 
the primitive of quantum mechanics. It may be mentioned that ``yes-no''-experiment depends primarily on the act of cognition.
In this framework it has been shown how the above kind of analysis and the concept of invariance will help us to 
understand the nature 
of ignorance (for example to understand the probabilistic nature of matching)  and hence the origin of probability in 
the context of brain function, similar to quantum physics, without using concepts like {\it collapse} or {\it measurement} as 
commonly used in quantum mechanics. It is curious to note that Edelman et al.(2002) pointed out that selection is biologically 
the more fundamental process. He conjectured that there exists two fundamental ways of patterning thought : {\it selectionism 
and logic}. We think that the selectionism play very sugnificant role in understanding information processing in brain.
\section{Discussion}
\noindent
The above analysis clearly indicates that it is not understandable how the anatomy of brain can permit the 
 joint space-time representation in the sense of special theory of relativity. So the applicability of any kind of quantum field theoretic 
approach  is not realizable, at least, at the present stage of understanding of brain function. However, 
it may be possible to define a smooth distance function and metric tensor 
in the probabilistic sense. Therefore, the probabilistic geometry seems to play a significant role in understanding Hilbert 
space structure and its connection to non-relativistic quantum mechanics. This approach sheds new light to undertstand the 
information processing and measurement procedurerelated to the brain. The implication of stochastic geometry in the inner world 
might have significant effects in the external world too, which will be considered in subsequent publications.

\acknowledgements
One of the authors (S.Roy) greatly acknowledges Center for Earth Observing and Space Research, School of computational 
Sciences, George Mason University, USA for their kind hospitality and funding for this work.

\end{document}